\documentstyle[preprint,aps,prl,multicol,epsfig]{revtex}

\begin{document}
\title{Nonextensive statistical mechanics: A brief introduction \thanks{Invited paper to appear in  {\it Extensive and non-extensive entropy and statistical mechanics},   a topical issue of {\it Continuum Mechanics and Thermodynamics (2003)}, edited by M. Sugiyama.} }
\author{Constantino Tsallis and Edgardo Brigatti \thanks{tsallis@cbpf.br, edgardo@cbpf.br}
}
\address{Centro Brasileiro de Pesquisas F\'{\i}sicas\\
Rua Xavier Sigaud 150, 
22290-180 Rio de Janeiro-RJ, Brazil.}
\maketitle

\begin{abstract}
Boltzmann-Gibbs statistical mechanics is based on the entropy $S_{BG}=-k \sum_{i=1}^W p_i \ln p_i$. It enables a successful thermal approach of ubiquitous systems, such as those involving short-range interactions, markovian processes, and, generally speaking, those systems whose dynamical occupancy of phase space tends to be ergodic. For systems whose microscopic dynamics is more complex, it is natural to expect that the dynamical occupancy of phase space will have a less trivial structure, for example a (multi)fractal or hierarchical geometry. The question naturally arises whether it is possible to study such systems with concepts and methods similar to those of standard statistical mechanics. The answer appears to be {\it yes} for ubiquitous systems, but the concept of entropy needs to be adequately generalized. Some classes of such systems can be satisfactorily approached with the entropy $S_q=k\frac{1-\sum_{i=1}^W p_i^q}{q-1}$ (with $q \in \cal R$, and $S_1 =S_{BG}$). This theory is sometimes referred in the literature as {\it nonextensive statistical mechanics}. We provide here a brief introduction to the formalism, its dynamical foundations, and some illustrative applications. In addition to these, we illustrate with a few examples the concept of {\it stability} (or {\it experimental robustness}) introduced by B. Lesche in 1982 and recently revisited by S. Abe.   
\end{abstract}



\section{INTRODUCTION}

Thermodynamics is based on two pillars: {\it energy} and {\it entropy}. The first one concerns (dynamical or mechanical) {\it possibilities}; the second one concerns the {\it probabilities} of those possibilities. The first one is more basic, and clearly depends on the {\it physical system} (classical, quantum, relativistic, or any other); the second one is more subtle,  and reflects the {\it information} upon the physical  system. It is obvious that the first one, the energy, is directly associated with the specific system, characterized for example by its Hamiltonian (which includes possible inertial terms, possible interactions, possible presence of external fields).   It is {\it much less} obvious that the same seems to happen with the second one, the entropy. It was long believed that the microscopic expression of the physical entropy {\it had} to be universal, i.e., system-independent (excepting for the trivial fact that it has to depend on $W$). More precisely, it {\it had} to be, for {\it all} systems, the well known one, namely
\begin{equation}
S_{BG}(\{p_i\})=-k \sum_{i=1}^W p_i \ln p_i \;,
\end{equation}
with its celebrated expression for equal probabilities
\begin{equation}
S_{BG}(p_i=1/W, \forall i)=k \ln W \;,
\end{equation}
$W$ being the total number of possibilities of the system under overall restrictions such as its total energy, total number of particles, and similar ones. 
However, this widespread belief of universality appears to have no rigorous basis. Indeed, it appears nowadays that the concept of physical information, and its microscopic expression in terms of probabilities, must be adapted to the system about which it is providing information, pretty much like impedances of interconnected circuits have to be similar in order to have an efficient and useful communication. Expressions (1) and (2) are so commonly used because most of the systems whose thermal properties are studied belong to the type involving ({\it strong}) chaos in its microscopic dynamics, i.e., {\it positive} Lyapunov exponents. There is no fundamental reason for which the {\it same} expression should necessarily be used for systems involving say a {\it vanishing} Lyapunov spectrum, i.e., for systems exhibiting {\it weak} chaos. Indeed, such systems, if isolated, might have serious difficulties in satisfying the ergodic hypothesis during the observational time of physical measures.

It is clear that the above statements about the nonuniversality of the microscopic expression for the entropy are {\it by no means} self-evident. It is through a variety of recent verifications that we have come to such possibility. The present paper focuses on some of this growing evidence. If, however, $S_{BG}$ is not universal, how to generalize it? 
No logical-deductive path ever existed for proposing a new physical theory, or for generalizing a pre-existing one. In fact, such proposal frequently --- perhaps always --- occurs on a metaphorical basis. 
The analysis of the structure of the BG theory provides us a metaphor for formulating a statistical mechanics which might be more powerful than the one we already have thanks to the genius of Boltzmann, Gibbs, and others.  It is perhaps worthy at this stage to explicitly insist  that we are talking of a {\it generalization} of the BG theory, {\it by no means of an alternative to it}. Without further delays, let us construct \cite{tsallis} a statistical mechanics  based on the following expression for the entropy:
\begin{equation}
S_q(\{p_i\})=k\frac{1-\sum_{i=1}^W p_i^q}{q-1}  \;\;\;\; ({q \in \cal R}; \; S_1 =S_{BG}) \;,
\end{equation}
with the following expression for equal probabilities
\begin{equation}
S_q(p_i=1/W, \forall i)=k \frac{W^{1-q}-1}{1-q} \equiv k \ln_q W  \;\; (\ln_1W=\ln W) . 
\end{equation}
If $A$ and $B$ are two independent systems (i.e., $p_{ij}^{A+B}=p_i^A p_j^B$), we verify that
\begin{equation}
\frac{S_q(A+B)}{k}=\frac{S_q(A)}{k}+\frac{S_q(B)}{k}+(1-q)\frac{S_q(A)}{k}\frac{S_q(B)}{k} \;,
\end{equation}
hence subextensivity (superextensivity) occurs if $q>1$ ($q<1$). Furthermore, it can be shown that the nonnegative entropy $S_q$ is concave (convex) for $q>0$ ($q<0$). 

Under appropriate canonical constraint (in addition to the trivial one $\sum_{i=1}^W p_i=1$), the entropy (3) is optimized (maximized for $q>0$, and minimized for $q<0$) by the following distribution \cite{tsallis,tsallis2,tsallis3}
\begin{equation}
p_i = \frac{e_q^{-\beta^\prime E_i}}{    \sum_{j=1}^W e_q^{-\beta^\prime E_j}    } \;,
\end{equation}
where $E_i$ is the energy of the $i-th$ microscopic state, $\beta^\prime$ plays the role of an inverse temperature, and the $q$-exponential function is defined as the inverse of the $\ln_q x$ function, i.e., $e_q^x \equiv [1+(1-q)x]^{\frac{1}{1-q}}$ ($e_1^x=e^x$).  A set of mini-reviews on the subject can be found in \cite{review}. The main properties of the theory, as well as its connection with thermodynamics, are there explained with great detail. 

\section{dynamical foundations}

It is clear that for the above theory to be complete we need to indicate how the entropic index $q$ can in principle be calculated {\it a priori} for a given system. Consistently with the ideas of Einstein \cite{einstein}, Krylov \cite{krylov}, Cohen \cite{cohen}, Baranger \cite{baranger}, and many others, the value of $q$ must be hidden in the microscopic (or mesoscopic) dynamics of the system. A large amount of examples enable to illustrate that it is indeed so. We shall restrict here to just one of them, a very simple one, namely the  family of logistic maps. Many others can be found in \cite{review}.  

Consider the one-dimensional dissipative map
\begin{equation}
x_{t+1}=1-a |x_t|^z \;\;\;(0 \le a \le2; \; z>1) \;,
\end{equation}
with $-1\le x_t \le 1$ and $t=0,1,2,..$. A value $a_c(z)$ exists for this map (e.g., $a_c(2)=1.401155...$) such that for $a>a_c(z)$ ($a<a_c(z)$) the Lyapunov exponent $\lambda_1$ tends to be positive (negative). At the edge of chaos, $a=a_c(z)$, the Lyapunov exponent precisely vanishes. For all values of $a$ such that the Lyapunov exponent exponent is different from zero, we have that the sensitivity to the initial conditions $\xi \equiv \lim_{\Delta x(0) \to 0} [\Delta x(t)/\Delta x(0)]=e^{\lambda_1 t}$. But at $a_c(z)$ we have $\xi = e_{q_{sen}}^{\lambda_{q_{sen}}t}$ \cite{TPZ,baldovinrobledo,baldovinrobledo2}, with $\lambda_{q_{sen}}(z)>0$, and $q_{sen}(z)$ varying from $-\infty$ to almost unity, when $z$ varies from unity to infinity ({\it sen} stands for {\it sensitivity}). For example 
\begin{equation}
q_{sen}(2)=0.2445... 
\end{equation}
This same value can be found through a connection with multifractal geometry. If we denote by $f(\alpha)$ the multifractal function, and $\alpha_{min}$ and $\alpha_{max}$ the two values of $\alpha$ at which $f(\alpha)$ vanishes, we can argue \cite{lyratsallis} that
\begin{equation}
\frac{1}{1-q_{sen}(z)}= \frac{1}{\alpha_{min}(z)} -  \frac{1}{\alpha_{max}(z)}= \frac{(z-1) \ln \alpha_F(z)}{\ln 2} \;,
\end{equation}
where $\alpha_F(z)$ is the $z$-generalization of the Feigenbaum universal constant.

The same value $q_{sen}(z)$ can be found through a third method, namely by studying the entropy production per unit time. We partition the $[-1,1]$ $x$ interval in $W$ little windows. We choose one of them and place (randomly or regularly) $M$ initial conditions inside it. We then follow those points as a function in time, and get the set of occupation numbers $\{M_i(t)\}$ ($ \sum_{i=1}^WM_i(t)=M$). We then define a probability set through $p_i(t) \equiv M_i(t)/M \;(\forall i)$, and calculate $S_q(t)/k$ using Eq. (3), and fixing some value for $q$. We then average over all initial windows (or a large number of them randomly chosen, which turns out to be numerically equivalent), and obtain $\langle S_q \rangle(t)/k$. We finally define the {\it entropy production per unit time} as follows
\begin{equation}
K_q \equiv \lim_{t\to \infty} \lim_{W\to \infty} \lim_{M\to \infty} \frac{\langle S_q \rangle(t)/k}{t} \;.
\end{equation} 
This concept essentially reproduces the so called Kolmogorov-Sinai entropy although generalized for arbitrary value of $q$. The equivalent of the usual Kolmogorov-Sinai entropy (which is fact not an entropy but an entropy production) is recovered for $q=1$. We can verify for the $z$-logistic map that $K_q$ vanishes (diverges) for $q>q_{sen}$ ($q<q_{sen}$), and that $K_{q{sen}}$ is {\it finite}, and coincident with $\lambda_{q_{sen}}$ ($q$-generalization of Pesin theorem \cite{TPZ,baldovinrobledo,baldovinrobledo2,LBRT}). 

It is finally possible to recover $q_{sen}$ through a forth method, namely through the {\it Lebesgue measure shrinking}: see \cite{moura,borges} for details.

Through this example of the logistic maps we have illustrated many relevant concepts concerning the index $q$ and the dynamical foundations of nonextensive statistical mechanics. One or other of these concepts, as well as a few somewhat different, have been shown for quantum chaos \cite{weinstein}, cercle maps \cite{ugurcercle}, Henon map \cite{tirnaklihenon}, standard maps \cite{brigatti,ananos}, lattice Lotka-Volterra model \cite{provata}, lattice Boltzmann model for the Navier-Stokes equations \cite{boghosian}, growth of free-scale networks \cite{barabasi} (the distribution turns out to be a $q$-exponential), long-range-interacting many-body Hamiltonian classical ferromagnets \cite{antoniruffo,celiaconstantino,giansanti,vitoandreaconstantino,aging,cabral,tsallisreply,moyano,nobre,ernestoising}, Lennard-Jones clusters \cite{doye} (the distributions are in fact $q$-exponentials), transport in quantum optical lattice \cite{lutz}, correlated \cite{bukman}, L\'evy-like \cite{levy} and other anomalous diffusions \cite{bologna,lenzimendestsallis}, multiplicative noise \cite{celialangevin} and dichotomic colored noise \cite{caceres} Langevin-like diffusions, among others.

\section{some applications}

The above theory has been applied in the literature to a large variety of systems, in addition to the ones we have just mentioned. Some of those attempts use various degrees of phenomenology, the $q$ index being obtained from fitting of experimental or computational available data. This is of course not an intrinsic necessity, but rather the consequence of the ignorance of the exact microscopic or mesoscopic dynamics of the natural or artificial system that is focused.   

Examples that have been analyzed include re-association in folded proteins \cite{bemski}, fluxes of cosmic rays \cite{cosmic},  turbulence \cite{beck,beckswinney,arimitsu}, finance and economics \cite{economics}, electron-positron annihilation \cite{bediaga}, motion of {\it Hydra} cells \cite{hydra}, epilepsy \cite{epilepsy},  linguistics \cite{linguistics}, nuclear physics \cite{creta}, astrophysics \cite{tsalliscordoba}, distributions in music \cite{music}, urban agglomerations \cite{urban}, internet phenomena \cite{internet},  among others. 

Finally, a special mention deserve the numerous applications that have been done \cite{simulated} in algorithms for global optimization (e.g., the {\it generalized simulated annealing}) and related computational methods.  

\section{stability or experimental robustness}

Lesche introduced \cite{lesche} two decades ago a quite interesting property that $S_{BG}$ satisfies. He named this property {\it stability}. However, in a recent private conversation, one of us (CT) proposed to him the use of the denomination {\it experimental robustness} instead. He agreed that such denomination reflected better the physical content of the mathematical property he had himself introduced. In addition to this, the expression ``stability" might be confused with ``thermodynamical stability", which has to do with the concavity of the entropy, a completely different and independent property. We shall therefore use either {\it stability} or {\it experimental robustness}  indistinctively. This property was further discussed recently by Abe \cite{abe,abepreprint}. In the present Section, we shall illustrate the property in a few typical cases, and will even start analyzing the possibility of enlarging it to metrics other than the one originally used in \cite{lesche}. The property consists in a specific kind of continuity. If two probability distributions are close to each other (which corresponds to slightly different realizations of a given experiment), this is expected to correspond to a relatively small discrepancy in any physical functional of the probability distribution, in particular in the entropy. 

The mathematical expression of this basic property follows. We first introduce a {\it distance} $d_\alpha \equiv ||p -p^\prime||_\alpha$ between two probability sets $\{p_i\}$ and $\{p_i^\prime\}$ associated with the same system:
\begin{equation}
d_\alpha \equiv \Bigl[\,\sum_{i=1}^W |p_i-p_i^\prime|^\alpha\,\Bigr]^{1/\alpha} \;\;\;\;(\alpha \ge 1)\,.
\end{equation}
Lesche used $\alpha=1$; we shall enlarge here his definition (e.g., $\alpha=2$ corresponds to the Pithagorean distance). The interval $0 <\alpha <1$ also defines a quantity which could be temptatively thought as measuring distances. However it does not constitute a metric, because it violates the triangular inequality. We shall therefore not consider that possibility at the present stage. 

We define now the {\it relative discrepancy} $R$ of the entropy as follows:
\begin{equation}
R \equiv \frac{S(\{p_i\}) -S(\{p_i^\prime\})}{S_{max}} \;,
\end{equation} 
where $S_{max}$ is the maximal value that the entropy under consideration can attain. For instance, the maximal value for $S_{BG}$ is $k \ln W$. 

We will say that $S$ is $\alpha$-stable (or experimentally $\alpha$-robust) if and only if, for any given $\epsilon>0$, a $\delta_\epsilon >0$ exists such that, independently from $W$,
\begin{equation}
d_\alpha \le \delta_\epsilon\; \Longrightarrow  \; |R|<\epsilon \;. 
\end{equation}
This implies, in particular, that $\lim_{d_\alpha \to 0} \lim_{W \to \infty} R(d_\alpha,W) = 0$ ($\lim_{W \to \infty} \lim_{d_\alpha \to 0} R(d_\alpha,W)$ always vanishes).

Let us apply definition (13) to the present nonextensive entropy $S_q$, to the Renyi entropy $S_q^R \equiv \ln \sum_{i=1}^W p_i^q/(q-1) = \ln [1+(1-q)S_q/k]/(1-q)$, to the normalized nonextensive entropy \cite{normalized}  $S_q^N \equiv  S_q/\sum_{i=1}^W p_i^q = S_q /[1+(1-q)S_q/k]$, and to the escort entropy $S_q^E$ defined as follows \cite{tsallis3}:
\begin{equation}
S_q^E(\{p_i\})=k\frac{1-(\sum_{j=1}^W p_j^{1/q})^{-q}}{q-1}
\end{equation}
This entropy emerges from $S_q(\{p_i\})$ where we do the escort transformation, namely $p_i=P_i^{1/q} /\sum_{j=1}^W P_j^{1/q}$, and then, for convenience, rename $P_i \to p_i$. 

Lesche showed \cite{lesche} that $S_{BG}$ is $1$-stable and that $S_q^R$ is not for any $q \ne 1$. Abe showed \cite{abe,abepreprint} that $S_q$ is $1$-stable for all $q>0$, whereas $S_q^N$ is not for any value of $q \ne 1$. We shall illustrate here these facts, including the case $\alpha>1$. In addition to this, we shall also illustrate that neither $S_q^E$ is $1$-stable. To do so we shall consider, following Abe, two typical cases, from now on referred to as {\it quasi-certainty} (QC) and as {\it quasi-equal-probabilities} (QEP). 

The quasi-certainty case refers to the following:
\begin{equation}
p_1=1; \; p_i=0 \; (\forall i \ne 1) \;,
\end{equation}
and
\begin{equation}
p_1^\prime=1-\frac{\delta}{2}; \; p_i^\prime=\frac{\delta}{2}\frac{1}{W-1} \; (\forall i \ne 1) \;,
\end{equation}
hence
\begin{equation}
d_\alpha= \Bigl(\frac{\delta}{2}\Bigr) \Bigl[1 + \frac{1}{(W-1)^{\alpha-1}}\Bigr]^{1/\alpha} \;,
\end{equation}
consequently
\begin{equation}
\delta=\frac{2 \,d_\alpha}{\Bigl[1 + \frac{1}{(W-1)^{\alpha-1}}\Bigr]^{1/\alpha}} \;.
\end{equation}
We verify that $d_1=\delta$ independs from $W$, whereas, for $\alpha>1$, $d_\alpha$ depends from $W$, and $\lim_{W\to\infty}d_\alpha=\delta/2$ ($\forall \alpha>1$).

The quasi-equal-probabilities case refer to the following:
\begin{equation}
p_1=0 ;\;p_i=\frac{1}{W-1}\;(\forall i \ne 1) \;,
\end{equation}
and
\begin{equation}
p_1^\prime=\frac{\delta}{2} ;\; p_i^\prime = \Bigl(1-\frac{\delta}{2}\Bigr)\frac{1}{W-1} \;,
\end{equation}
hence
\begin{equation}
d_\alpha= \frac{\delta}{2} \Bigl[1 + \frac{1}{(W-1)^{\alpha-1}} \Bigr]^{1/\alpha} \;,
\end{equation}
consequently
\begin{equation}
\delta=\frac{2 \,d_\alpha}{\Bigl[1 + \frac{1}{(W-1)^{\alpha-1}}\Bigr]^{1/\alpha}} \;.
\end{equation}
Once again we verify that $d_1=\delta$ independs from $W$, whereas, for $\alpha>1$, $d_\alpha$ depends from $W$, and $\lim_{W\to\infty}d_\alpha=\delta/2$ ($\forall \alpha>1$). Notice that, for this QEP case, the results in Eqs. (21) and (22) turn out to exactly coincide with those corresponding to the QC case, namely Eqs. (17) and (18).

We may now calculate, as functions of $(\delta,W,\alpha,q)$ and for both quasi-certainty and quasi-equal-probabilities cases, the quantities $R_q$, $R_q^R$, $R_q^N$ and $R_q^E$, respectively associated with the entropic functionals $S_q$, $S_q^R$, $S_q^N$ and $S_q^E$. We obtain the following results:
\begin{equation}
R_q=-\frac{(1-\delta/2)^{q}+(\delta/2)^{q}(W-1)^{1-q}-1}{W^{1-q}-1}~~~(QC)\nonumber \\
\end{equation}
\begin{equation}
R_q=\frac{(W-1)^{1-q}-(\delta/2)^{q}-(1-\delta/2)^{q}(W-1)^{1-q}}{W^{1-q}-1}~~~(QEP)\nonumber\\
\end{equation}
\begin{equation}
R_q^R=  -\frac{ \ln{[(1-\delta/2)^{q}+(\delta/2)^{q}(W-1)^{1-q}}]}{(1-q) \ln W }~~(QC)\nonumber\\
\end{equation}
\begin{equation}
R_q^R=  \frac{(1-q) \ln{(W-1)}-\ln{[(\delta/2)^{q}+(1-\delta/2)^{q}(W-1)^{1-q}}]}{(1-q)\ln W }~~(QEP)\nonumber\\
\end{equation}
\begin{equation}
R_q^N=-\frac{1-1/[(1-\delta/2)^{q}+(\delta/2)^{q}(W-1)^{1-q} ]}{1-W^{q-1}}~~~(QC)\nonumber\\
\end{equation}
\begin{equation}
R_q^N=\frac{1/[(\delta/2)^{q}+(1-\delta/2)^{q}(W-1)^{1-q}]-(W-1)^{q-1}}{1-W^{q-1}}~~~(QEP)\nonumber \\
\end{equation}
\begin{equation}
R_q^E=-\frac{1-[(1-\delta/2)^{1/q}+(W-1)(\delta/2)^{1/q}(W-1)^{-1/q}]^{-q}}{W^{1-q}-1}~~~(QC) \nonumber\\
\end{equation}
\begin{equation}
R_q^E=\frac{ (W-1)^{1-q}-[(\delta/2)^{1/q}+(W-1)(1-\delta/2)^{1/q}(W-1)^{-1/q}]^{-q}}{W^{1-q}-1}~~~(QEP)\nonumber\\
\end{equation}
All these ratios can be expressed as explicit functions of $(d_\alpha, W,\alpha,q)$ if we use Eqs. (18, 22).

We present typical examples in Fig. 1 for $S_{BG}$, and in Figs. 2 and 3 for its four generalizations. We notice in Figs. 2 and 3 that {\it only} $R_q$ satisfies (13) for {\it all} the examples; $R_q^R$ violates (13) for QC with $q<1$, and for QEP with $q>1$; $R_q^N$ violates (13) for QC with $q<1$, and for QEP with $q>1$; 
$R_q^E$ violates (13) for QEP with $q<1$, and for QC with $q>1$.

At this point, let us make a remark about the $\alpha>1$ metrics. This class of distance satisfies, as already mentioned, the triangular inequality. This is why we have considered it until now. However, it presents an inconvenient peculiarity. Let us illustrate this through the following example (assuming $W$ even):
\begin{equation}
p_i=0 \;(i=1,2,...,\frac{W}{2}) ;\;p_i=\frac{2}{W}\;(i= \frac{W}{2}+1, \frac{W}{2}+2,  ...,W) \;,
\end{equation}
and
\begin{equation}
p_1^\prime=\frac{\delta}{2} \frac{2}{W}  \;(i=1,2,...,\frac{W}{2})   ;\; p_i^\prime = \Bigl(1-\frac{\delta}{2}\Bigr)  \frac{2}{W}  \;(i= \frac{W}{2}+1, \frac{W}{2}+2,  ...,W)  \;,
\end{equation}
hence
\begin{equation}
d_\alpha= \delta W^{(1-\alpha)/\alpha} \;.
\end{equation}
As for the QC and QEP cases considered above, $d_1=\delta$. {\it But}, for $\alpha>1$ and $W \to\infty$, $d_\alpha \to 0$. Consequently, although $\{p_i\} \ne \{p_i^\prime\}$, their distance $d_\alpha$ vanishes, a quite inconvenient feature.  We thus verify that $d_1$ is very special indeed, since it does not depend on $W$.

Finally, in order that it becomes clear that stability is a property  independent from concavity, we represent, in Fig. 4, typical cases (namely those associated with the $W=2$ system) of all four generalizations of $S_{BG}$ considered here. We notice that {\it only} $S_q$ satisfies concavity for {\it all} values of $q>0$; $S_q^R$ and $S_q^N$ are concave for $0 <q \le 1$, but not for $q>1$; $S_q^E$ is concave for $q \ge 1$, but not for $0<q<1$.

\section{conclusion}

We have provided a brief introduction to nonextensive statistical mechanics and to its connections to thermodynamics, as well as a few explanations about its dynamical foundations. Various applications have been mentioned, and the reader is referred to the appropriate references for further details. Many interesting open questions remain to be further investigated. Among them, let us mention the subtle, and yet not completely clarified, connection between the present ideas and long-range-interacting Hamiltonians.

In Section IV we have addressed the property of stability (or experimental robustness) introduced by Lesche in 1982. We have illustrated, for typical cases and metrics, the generic validity of this property for the nonextensive entropy $S_q$, as well as its nonvalidity for the (extensive) Renyi entropy $S_q^R$, for the normalized nonextensive entropy $S_q^N$, and for the escort nonextensive entropy $S_q^E$. In addition to this, we have also illustrated the validity or nonvalidity of the concavity property for the same entropic forms. 

Last but not least, the history of sciences teaches us that all attempts of innovation in the scientific theories have been the subject of objections or critical comments. The present attempt is no exception. We may mention the critique in \cite{luzzi}, counterbalanced in \cite{science1,science2,science3}, the critique in \cite{nauenberg}, counterbalanced in \cite{tsallisreply,moyano}, and the critique in \cite{zanettecritique}, counterbalanced in \cite{bologna,economics}.  

\section{acknowledgements}

We acknowledge very useful discussions with A.M.C. de Souza. This work has been partially supported by PRONEX, CNPq, FAPERJ and CAPES (Brazilian agencies).

\begin{figure}
\begin{center}
\includegraphics[width=12cm,angle=0]{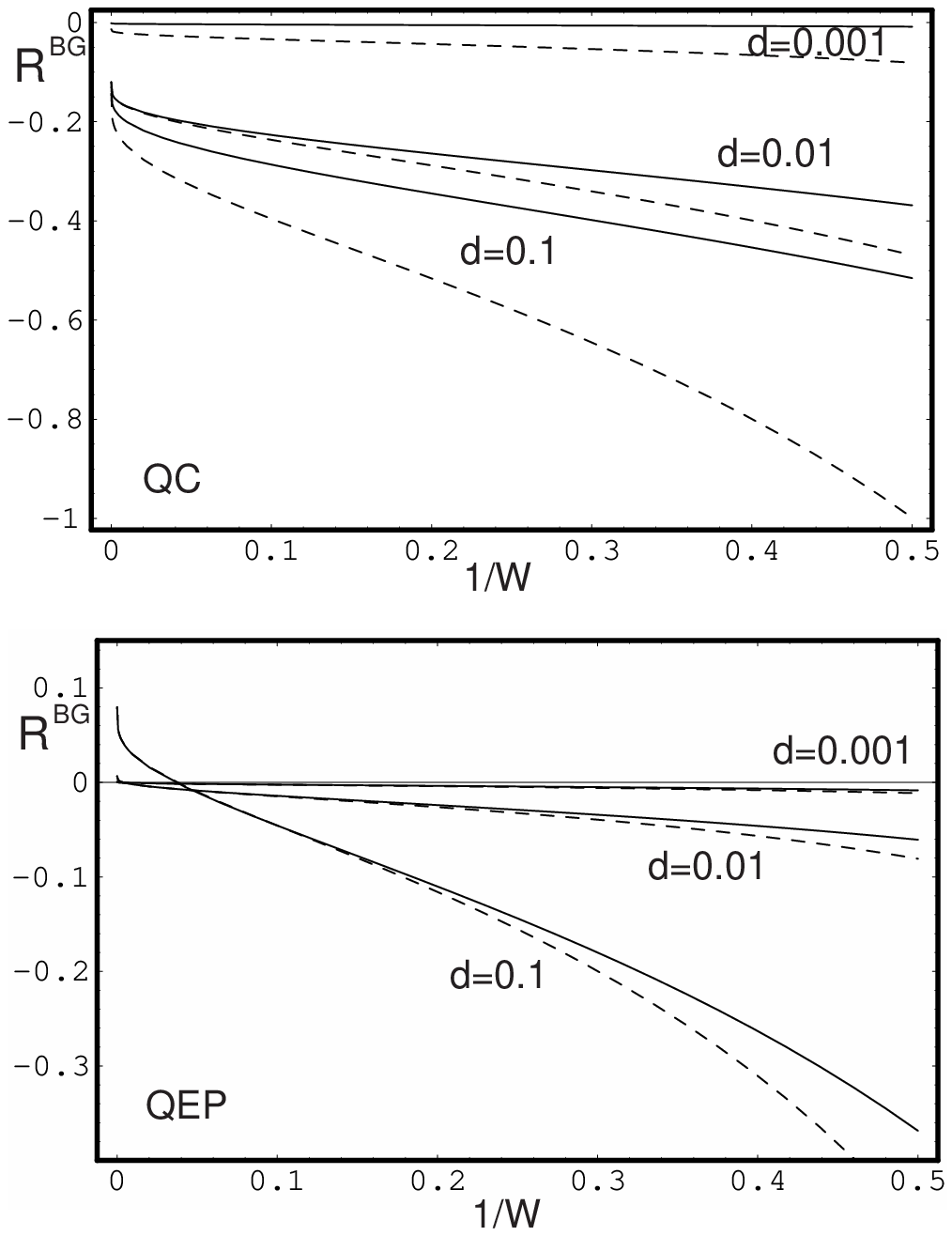}
\end{center}
\caption{\small Typical examples (QEP and QC) for the stability of the  $S_{BG}$.The dashed (continuous) curves correspond to the $\alpha = 1$ ($\alpha = 2$) metric.} 
\end{figure}

\newpage

\begin{figure}
\begin{center}
\includegraphics[width=15cm,angle=0]{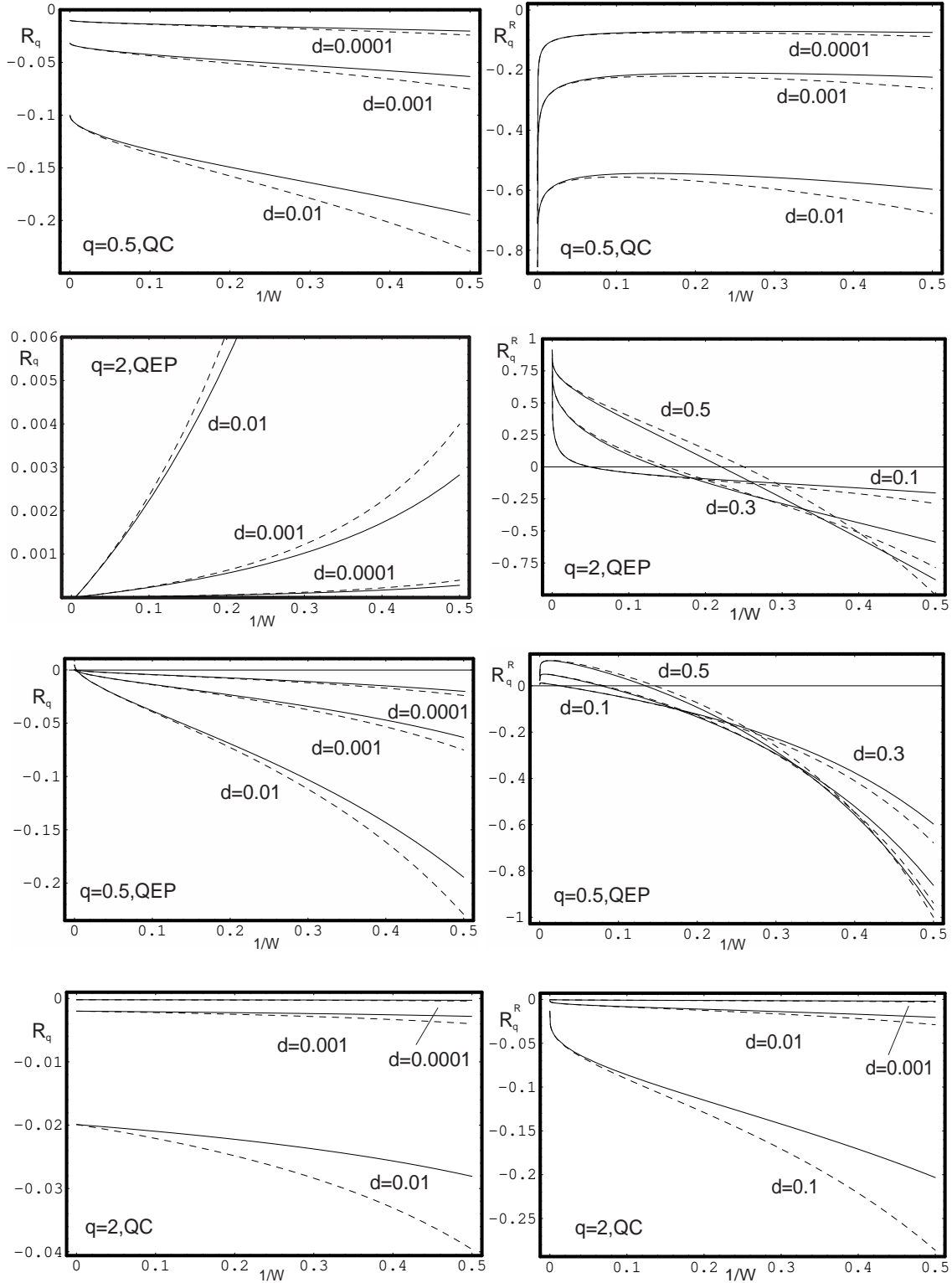}
\end{center}
\caption{\small  Typical examples (QEP and QC, $q<1$ and $q>1$) for the stability (or lack of stability) of the $S_q$ and $S_q^R$ entropies.  The dashed (continuous) curves correspond to the $\alpha = 1$ ($\alpha = 2$) metric.} 
\end{figure}

\begin{figure}
\begin{center}
\includegraphics[width=15cm,angle=0]{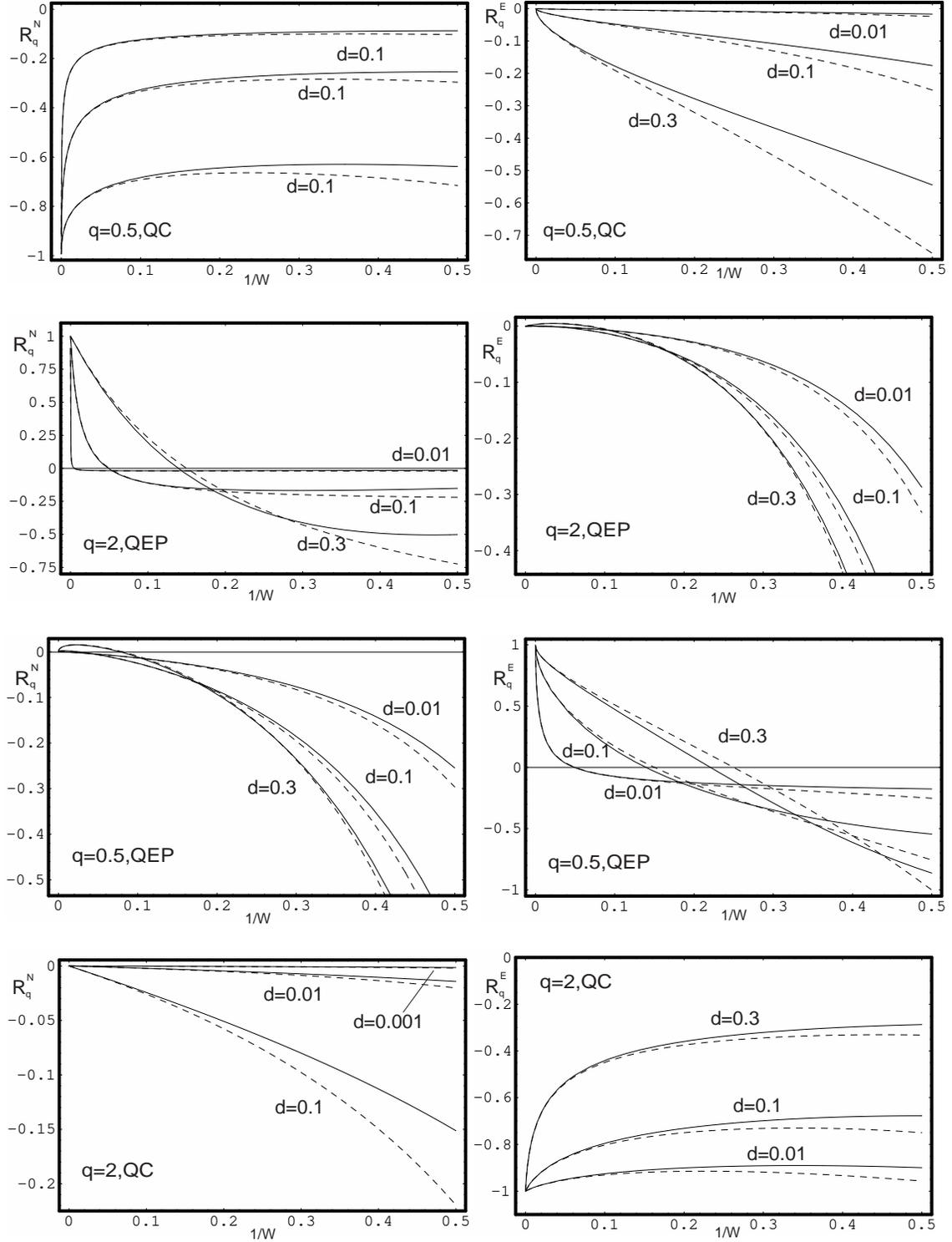}
\end{center}
\caption{\small The same as in Fig. 3, for the $S_q^N$ and $S_q^E$ entropies.} 
\end{figure}

\begin{figure}
\begin{center}
\includegraphics[width=17cm,angle=0]{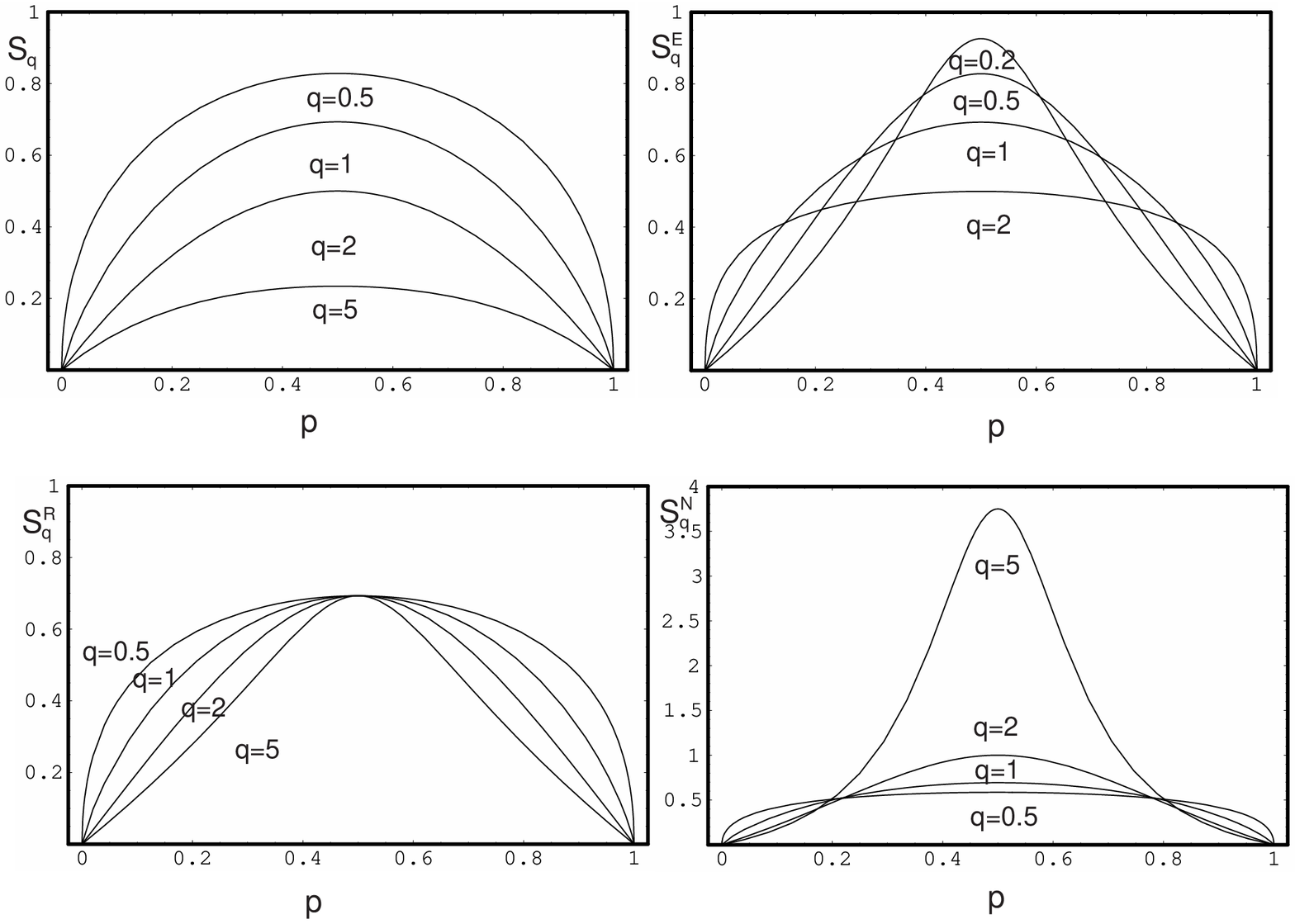}
\end{center}
\caption{\small  Typical examples exhibiting the concavity (or the lack of concavity) of the four entropies discussed here.} 
\end{figure}



\begin{thebibliography}{99}

\bibitem{tsallis} C. Tsallis, J. Stat. Phys. {\bf 52}, 479 (1988). 

\bibitem{tsallis2}E.M.F. Curado and C. Tsallis, J. Phys. A {\bf 24}, L69 (1991) [Corrigenda: {\bf 24}, 3187 (1991) and {\bf 25}, 1019 (1992)].

\bibitem{tsallis3}C. Tsallis, R.S. Mendes and A.R. Plastino, Physica A {\bf 261}, 534 (1998).

\bibitem{review} S.R.A. Salinas and C. Tsallis, eds., {\it Nonextensive Statistical Mechanics and Thermodynamics}, Brazilian Journal of Physics {\bf 29} (1999); S. Abe and Y. Okamoto, eds., {\it Nonextensive Statistical Mechanics and Its Applications, Series Lecture Notes in Physics} (Springer-Verlag, Berlin, 2001); G. Kaniadakis, M. Lissia and A. Rapisarda, eds., {\it Non Extensive Thermodynamics and Physical Applications}, Physica A {\bf 305} (Elsevier, Amsterdam, 2002); P. Grigolini, C. Tsallis and B.J. West, eds., {\it Classical and Quantum Complexity and Nonextensive Thermodynamics}, Chaos , Solitons and Fractals {\bf 13}, Number 3 (Pergamon-Elsevier, Amsterdam, 2002); 
M. Gell-Mann and C. Tsallis, eds. {\it Nonextensive Entropy -- Interdisciplinary Applications}, (Oxford University Press, Oxford, 2003), in press; H.L. Swinney and C. Tsallis, {\it Anomalous Distributions, Nonlinear Dynamics, and
Nonextensivity}, Physica D (Elsevier, Amsterdam, 2003), in press. 
A regularly updated bibliography on the subject can be found at http://tsallis.cat.cbpf.br/biblio.htm.

\bibitem{einstein} A. Einstein, Annalen der Physik {\bf 33}, 1275 (1910) [ {\it ``Usually $W$ is put equal to the number of complexions... In order to calculate $W$, one needs a complete (molecular-mechanical) theory of the system under consideration. Therefore it is dubious whether the Boltzmann principle has any meaning without a complete molecular-mechanical theory or some other theory which describes the elementary processes. 
$S=\frac{R}{N}\log W +$ const. seems without content, from a phenomenological point of view, without giving in addition such an Elementartheoráe.''} (Translation: Abraham Pais, Subtle is the Lord .... Oxford University Press, 1982)].

\bibitem{krylov} N.S. Krylov, Nature {\bf 153}, 709 (1944); N.S. Krylov, Works on the Foundations of Statistical Physics, translated by A.B. Migdal, Ya. G. Sinai and Yu. L. Zeeman, Princeton Series in Physics (Princeton University Press, Princeton, 1979).

\bibitem{cohen} E.G.D. Cohen, Physica A {\bf 305}, 19 (2002).

\bibitem{baranger} M. Baranger, Physica A {\bf 305}, 27 (2002).

\bibitem{TPZ} C. Tsallis, A.R. Plastino, and W.M. Zheng, Chaos, Solitons and Fractals {\bf 8}, 885 (1997).

\bibitem{baldovinrobledo} F. Baldovin and A. Robledo, Europhys. Lett. {\bf 60}, 518 (2002), and Phys. Rev. E {\bf 66}, 8045104 (2002).

\bibitem{baldovinrobledo2}F. Baldovin and A. Robledo, cond-mat/0304410 (2003).

\bibitem{lyratsallis} M.L. Lyra and C. Tsallis, Phys. Rev. Lett. 80, 53 (1998). 

\bibitem{LBRT} V. Latora, M. Baranger, A. Rapisarda, and C. Tsallis, Phys. Lett. A, 
{\bf 273} 97 (2000).

\bibitem{moura}F.A.B.F. de Moura, U. Tirnakli and M.L. Lyra, Phys. Rev. E {\bf 62}, 6361 (2000).

\bibitem{borges}E.P. Borges, C. Tsallis, G.F.J. Ananos and P.M.C. Oliveira, Phys. Rev. Lett. {\bf 89}, 254103 (2002).

\bibitem{weinstein} Y.S. Weinstein, S. Lloyd and C. Tsallis, Phys. Rev. Lett. {\bf 89}, 214101 (2002); Y.S Weinstein, C. Tsallis and S. Lloyd, {\it On the emergence of nonextensivity at the edge of quantum chaos}, in {\it Decoherence and Entropy in Complex Systems}, ed. H.T. Elze, Lecture Notes in Physics (Springer, Heidelberg, 2003), in press.

\bibitem{ugurcercle}U. Tirnakli, C. Tsallis and M.L. Lyra,  Phys. Rev. E {\bf 65}, 036207 (2002).

\bibitem{tirnaklihenon}U. Tirnakli, Phys. Rev. E {\bf 66}, 066212 (2002); E.P. Borges and U. Tirnakli, cond-mat/0302616.

\bibitem{brigatti}F. Baldovin, E. Brigatti and C. Tsallis, cond-mat/0302559. 

\bibitem{ananos}G.F.J. Ananos, F. Baldovin and C. Tsallis, {\it Nonstandard sensitivity to initial conditions and entropy production in standard maps}, preprint (2003).

\bibitem{provata}G.A. Tsekouras, A. Provata and C. Tsallis, cond-mat/0303104.

\bibitem{boghosian}B.M. Boghosian, P.J. Love, P.V. Coveney, I.V. Karlin, S. Succi and J. Yepez, cond-mat/0211093;  
B.M. Boghosian, P. Love and J. Yepez, {\it Galilean-invariant multi-speed entropic lattice Boltzmann models}, preprint (2003)

\bibitem{barabasi} R. Albert and A.L. Barabasi, Phys. Rev. Lett. {\bf 85}, 5234 (2000).

\bibitem{antoniruffo}M. Antoni and S. Ruffo, Phys. Rev. E {\bf 52}, 2361 (1995).

\bibitem{celiaconstantino}C. Anteneodo and C. Tsallis, Phys. Rev. Lett. {\bf 80}, 5313 (1998).

\bibitem{giansanti}A. Campa, A. Giansanti, D. Moroni and C. Tsallis, Phys. Lett. A {\bf 286}, 251 (2001).

\bibitem{vitoandreaconstantino}V. Latora, A. Rapisarda and C. Tsallis, Phys. Rev. E {\bf 64}, 056134 (2001).

\bibitem{aging}M.A. Montemurro, F. Tamarit and C. Anteneodo, Phys. Rev. E {\bf 67}, 031106 (2003).

\bibitem{cabral}B.J.C. Cabral and C. Tsallis, Phys. Rev. E {\bf 66}, 065101(R) (2002).

\bibitem{tsallisreply}C. Tsallis, cond-mat/0304696 (2003).

\bibitem{moyano}L.G. Moyano, F. Baldovin and C. Tsallis, cond-mat/0305091. 

\bibitem{nobre}F.D. Nobre and C. Tsallis, cond-mat/0301492.

\bibitem{ernestoising}E.P. Borges, C. Tsallis, A. Giansanti and D. Moroni, to appear in a volume honoring S.R.A. Salinas (2003) [in Portuguese].

\bibitem{doye}J.P.K. Doye, Phys. Rev. Lett. {\bf 88}, 238701 (2002).

\bibitem{lutz}E. Lutz, Phys. Rev. A  (2003), in press [cond-mat/0210022].

\bibitem{bukman}A.R. Plastino and A. Plastino, Physica A  {\bf 222}, 347 (1995); C. Tsallis and D.J. Bukman, Phys. Rev. E {\bf 54}, R2197 (1996); E.M.F. Curado and F.D. Nobre, Phys. Rev. E {\bf 67}, 021107 (2003).

\bibitem{levy}P.A. Alemany and D.H. Zanette, Phys. Rev. E {\bf 49}, R956 (1994); C. Tsallis, S.V.F Levy, A.M.C. de Souza and R. Maynard,, Phys. Rev. Lett. {\bf 75}, 3589 (1995) [Erratum: {\bf 77}, 5442 (1996)]; D. Prato and C. Tsallis, Phys. Rev. E {\bf 60}, 2398 (1999).

\bibitem{bologna}M. Bologna, C. Tsallis and P. Grigolini, Phys. Rev. E {\bf 62}, 2213 (2000).

\bibitem{lenzimendestsallis}C. Tsallis and E.K. Lenzi, in {\it Strange Kinetics}, eds. R. Hilfer et al, Chem. Phys.  {\bf 284}, 341 (2002) [Erratum: {\bf 287}, 341 (2002)]; E.K. Lenzi, R.S. Mendes and C. Tsallis, Phys. Rev. E {\bf 67}, 031104 (2003).

\bibitem{celialangevin}C. Anteneodo and C. Tsallis, cond-mat/0205314.

\bibitem{caceres}M.O. Caceres, Phys. Rev. E {\bf67}, 016102 (2003). 

\bibitem{bemski}C. Tsallis, G. Bemski and R.S. Mendes, Phys. Lett. A {\bf 257}, 93 (1999).

\bibitem{cosmic}C. Tsallis, J.C. Anjos and E.P. Borges, Phys. Lett. A {\bf 310}, 372  (2003).

\bibitem{beck}C. Beck, Phys. Rev. Lett. {\bf 87}, 180601 (2001).

\bibitem{beckswinney}C. Beck, G.S. Lewis and H.L. Swinney, Phys. Rev. E {\bf 63}, 035303 (2001).

\bibitem{arimitsu}N. Arimitsu and T. Arimitsu, Europhys. Lett. {\bf 60}, 60 (2002).

\bibitem{economics}L. Borland, Phys. Rev. Lett. {\bf 89}, 098701 (2002); C. Tsallis, C. Anteneodo, L. Borland and R. Osorio, Physica A (2003), in press [cond-mat/0301307]. 

\bibitem{bediaga}I. Bediaga, E.M.F. Curado and J. Miranda, Physica A {\bf 286}, 156 (2000).

\bibitem{hydra}A. Upadhyaya, J.-P. Rieu, J.A. Glazier and Y. Sawada, Physica A {\bf 293}, 549 (2001).

\bibitem{epilepsy}O.A. Rosso, M.T. Martin and A. Plastino, Physica A {\bf 313}, 587 (2002).

\bibitem{linguistics}M.A. Montemurro, Physica A {\bf 300}, 567 (2001).

\bibitem{creta}C. Tsallis and E.P. Borges, {\it Nonextensive statistical mechanics - Applications to nuclear and high energy physics}, Proc. {\it Xth International Workshop on Multiparticle Production - Correlations and Fluctuations in QCD}, ed. N. Antoniou (World Scientific, Singapore, 2003), to appear [cond-mat/0301521].

\bibitem{tsalliscordoba}C. Tsallis, D. Prato and A.R. Plastino, {\it Nonextensive statistical mechanics: Some links with astronomical phenomena}, to appear in the Proceedings of the {\it XIth United Nations / European Space Agency Workshop on Basic Space Sciences}, Office for Outer Space Affairs / United Nations (Cordoba, 9-13 September 2002), ed. H. Haubold, special issue of Astrophysics and Space Science (Kluwer Academic Publishers, Dordrecht, 2003). 

\bibitem{music}E.P. Borges, Eur. Phys. J. B  {\bf 30}, 593 (2002).

\bibitem{urban}L.C. Malacarne, R.S. Mendes and E.K. Lenzi, Phys. Rev. E {\bf 65}, 017106 (2002).

\bibitem{internet}S. Abe and N. Suzuki, Phys. Rev. E {\bf 67}, 016106 (2003).

\bibitem{simulated}C. Tsallis and D.A. Stariolo, Physica A {\bf 233}, 395 (1996); a preliminary version appeared (in English) as Notas de Fisica/CBPF 026 (June 1994); T.J.P. Penna, Phys. Rev. E {\bf 51}, R1 (1995); K.C. Mundim and C. Tsallis, Int. J. Quantum Chem. {\bf 58}, 373 (1996); I. Andricioaei and J.E. Straub, Phys. Rev. E {\bf 53}, R3055 (1996); M.A. Moret, P.G. Pascutti, P.M. Bisch and K.C. Mundim, J. Comp. Chemistry {\bf 19}, 647 (1998); L. Guo, D.E. Ellis and K.C. Mundim, Journal of Porphyrins and Phthalocyanines {\bf 3}, 196 (1999); M.A. Moret, P.M. Bisch and F.M.C. Vieira, Phys. Rev. E {\bf 57}, R2535 (1998); A. Berner, D. Fuks, D.E. Ellis, K.C. Mundim and S. Dorfman, Applied Surface Science {\bf 144-145}, 677 (1999); P. Serra, A.F. Stanton, S. Kais and R.E. Bleil, J. Chem. Phys. {\bf 106}, 7170 (1997); Y. Xiang, D.Y. Sun, W. Fan and X.G. Gong, Phys. Lett. A {\bf 233}, 216 (1997); I. Andricioaei and J.E. Straub, J. Chem. Phys. {\bf 107}, 9117 (1997); U.H.E. Hansmann , M. Masuya and Y. Okamoto, Proc. Natl. Acad. Sci. USA {\bf 94}, 10652 (1997); M.R. Lemes, C.R. Zacharias and A. Dal Pino Jr., Phys. Rev. B {\bf 56}, 9279 (1997); Z.X. Yu and D. Mo, Thin Solid Films {\bf 425}, 108 (2003).

\bibitem{lesche} B. Lesche, J. Stat. Phys. {\bf 27}, 419 (1982).

\bibitem{abe} S. Abe, Phys. Rev. E {\bf 66}, 046134 (2002).

\bibitem{abepreprint}S. Abe, to appear in Continuum Mechanics and Thermodynamics (2003) [cond-mat/0305087]. 

\bibitem{normalized}P.T. Landsberg and V. Vedral, Phys. Lett. A {\bf 247}, 211 (1998); A.K. Rajagopal and S. Abe, Phys. Rev. Lett. 83, 1711 (1999). 

\bibitem{luzzi}R. Luzzi, A.R Vasconcellos and J. Galvao Ramos, Science {\bf 298}, 1171 (2002).

\bibitem{science1}S. Abe and A.K. Rajagopal, Science {\bf 300}, 249  (2003).

\bibitem{science2}A. Plastino,  Science {\bf 300}, 250  (2003).

\bibitem{science3}V. Latora, A. Rapisarda and A. Robledo, Science  {\bf 300}, 250  (2003).

\bibitem{nauenberg}M. Nauenberg, Phys. Rev. E {\bf 67}, 036114 (2003).

\bibitem{zanettecritique}D.H. Zanette and M.A. Montemurro,  cond-mat/0212327 (2002). 

\end{thebibliography}
\end{document}